\documentclass[aps,pra,twocolumn,superscriptaddress,longbibliography,floatfix,amsmath,amssymb%reprint,
]{revtex4-2}

\usepackage[T1]{fontenc}
\usepackage[utf8]{inputenc}
\usepackage{amsmath,amssymb,amsfonts,amsthm,mathtools}
\usepackage{physics}
\usepackage{xcolor}
\usepackage{siunitx}

\usepackage{graphicx}% Include figure files
\usepackage{dcolumn}% Align table columns on decimal point
\usepackage{bm}% bold math
\usepackage{hyperref}% add hypertext capabilities
\hypersetup{
  colorlinks=true,
  linkcolor=blue,
  citecolor=blue,
  urlcolor=blue
}

\theoremstyle{definition}

\definecolor{editgreen}{rgb}{0.0,0.45,0.0}
\definecolor{editred}{rgb}{0.75,0.0,0.0}
\definecolor{corrblue}{rgb}{0.0,0.0,0.75}

\begin{document}

% \preprint{APS/123-QED}

\title{Quantum enhanced beam displacement estimation: sensitivity limits, optimal probes, and measurement schemes}% Force line breaks with \\

\author{Balázs Novák}
\email{balazs.novak@uv.es}
\affiliation{Instituto de Física Corpuscular (IFIC), CSIC–Universitat de València, and Departament de Física Teòrica, UV, Avinguda Vicent Andrés Estellés, 19, E-46100 Burjassot (València), Spain}

\author{Pau Colomer}
\affiliation{Chair of Theoretical Information Technology, School of Computation, Information and Technology, Technical University of Munich, Theresienstra{\ss}e 90, 80333 M\"unchen, Germany}

\author{Joseph Zhang}
\affiliation{Department of Physics, Imperial College London, Prince Consort Road, SW7 2AZ London, UK}

\author{Emre Köse}
\email{emrekose@ific.uv.es}
\affiliation{Instituto de Física Corpuscular (IFIC), CSIC–Universitat de València, and Departament de Física Teòrica, UV, Avinguda Vicent Andrés Estellés, 19, E-46100 Burjassot (València), Spain}

\author{Manuel Gessner}
\email{manuel.gessner@uv.es}
\affiliation{Instituto de Física Corpuscular (IFIC), CSIC–Universitat de València, and Departament de Física Teòrica, UV, Avinguda Vicent Andrés Estellés, 19, E-46100 Burjassot (València), Spain}

\date{\today}% It is always \today, today,

\begin{abstract}
    We investigate the ultimate sensitivity limits for measuring small transverse beam displacements and identify optimal strategies that attain them, using the framework of mode parameter estimation. We contrast two families of states with different resource constraints, using Hermite-Gauss and Laguerre-Gauss mode pairs. First, for fixed total photon number states, the optimal state is identified and we show that it asymptotically approaches a rotated NOON state. The sensitivity bound is compared to that of an asymptotically optimal pure Gaussian state with fixed average photon number. Finally, we propose concrete measurement schemes using the method of moments and show that the twin-Fock state may provide a viable alternative to the rotated NOON state for this task. Our work provides a broader benchmark for developing practical strategies for spatial sensing at the quantum limit.
\end{abstract}

%Use showkeys class option if keyword
\maketitle

%\tableofcontents

\section{Introduction}

Precise measurement of transverse optical beam displacements is important in optical metrology, sensing, and alignment. Small changes in the position of an atomic force microscope's probe produce shifts of a reflected beam on a position-sensitive detector, providing a readout for surface imaging and force measurements \cite{Fontana2024,Wu:21}. Similar beam-deflection measurements are used in photothermal absorption spectroscopy \cite{Boccara1980}, while beam-position sensing enables tracking in free-space optical communication links \cite{Patel:24} and transmitter–receiver alignment in lidar \cite{Palm2021}. Spatial optical measurements are also used for pointing control in coronagraphic imaging in telescopes~\cite{Guyon:09} and cavity alignment in interferometry relevant to gravitational-wave detection \cite{Tao:24}. These applications motivate the investigation of quantum precision limits for beam-displacement sensing \cite{Fabre:00,Treps2005,Delaubert:06}. Related spatial-mode techniques also enable sub-Rayleigh estimation of the separation between incoherent optical sources \cite{Tsang:16,Boucher2020}.

Quantum metrology provides the tools that allow us to investigate ultimate limits of the precision attainable in estimating a parameter of interest \cite{Helstrom:76,Holevo:82,Caves:81,Giovannetti:06,Paris:09,PezzeSmerzi:14,Demkowicz:15}. Exploiting quantum properties has the potential to improve the absolute error of the estimator up to a scaling with the inverse of the fixed number of photons being measured, $1/N$, known as the Heisenberg limit. In contrast, classical states of light, and more generally particle-separable states of $N$ photons, are limited to an error scaling with the inverse of the square root of the photon number, $1/\sqrt{N}$, known as the standard quantum limit (SQL) \cite{BraunsteinCaves1994, PezzeSmerziPRL2009,Demkowicz2012,Toth_2014}. When the number of photons is allowed to fluctuate, the asymptotic sensitivity limit is not constrained by a fixed average photon number $\Bar{N}$.

Quantum enhancement beyond the SQL has already been demonstrated experimentally for the estimation of a beam displacement~\cite{Treps2003, Delaubert:06}. It was recognised that a single-mode approach that consists in populating exclusively the fundamental light mode of the beam of interest would never enable any enhancement over the SQL even if prepared in arbitrary nonclassical states~\cite{Fabre:00}. Instead, specific auxiliary modes are needed to enable quantum enhancements and must be prepared in nonclassical states with reduced quantum fluctuations below the SQL. Recent theoretical work~\cite{Gessner:23} has revealed the importance of optical mode shapes in such estimation problems: essential information about an unknown parameter can be carried by modes that were initially unpopulated.

In this article, we apply the framework of mode parameter estimation to transverse beam displacement in the Hermite--Gauss and Laguerre--Gauss mode bases, establishing the ultimate quantum limits on asymptotic precision and identifying optimal probe states and mode designs that attain them. We focus on two families of initial states with operationally distinct resource constraints. Firstly, we consider states that have a fixed total photon number. Such states arise naturally in heralded photonic experiments \cite{Stasi2023} and are naturally suited for direct photon counting without an external optical phase reference. When the receiver is based on photon-number-resolving or coincidence detection rather than phase-sensitive homodyne detection, fixed-number photonic probes provide a natural description of both the input state and the measurement outcomes. Such architectures have demonstrated unconditional sub-shot-noise sensing and loss-tolerant multiphoton metrology \cite{Slussarenko2017,Matthews2016}. We also consider optimal Gaussian states at fixed mean photon number as benchmarks. Their experimental relevance and accessibility have motivated extensive studies of Gaussian probes for mode-parameter estimation and spatial sensing \cite{Pinel2012,Sorelli:21,He:26,Reichert:26}, and our optimisation recovers these results in this regime.

This article is structured as follows: In Sec.~\ref{sec: framework}, we summarise the mode parameter estimation framework for estimating transverse beam-displacement magnitude in the Hermite--Gauss and Laguerre--Gauss bases. We recover the known results that single-mode beams cannot provide quantum enhancement and that suitable mode pairs enable Heisenberg scaling for small displacements. In Sec.~\ref{sec: limits}, we derive the ultimate two-mode precision limit at fixed total photon number and compare it with the sensitivity of the asymptotically optimal centred pure two-mode Gaussian state, illustrating the results with simple examples. Finally, in Sec.~\ref{sec: Measurement schemes}, we propose practical measurement schemes based on the method of moments for fixed-photon-number states and a squeezed Gaussian probe, evaluate their moment-based sensitivities, and demonstrate quantum advantage.

\section{Framework}\label{sec: framework}

Suppose a photon beam is radially displaced in the transverse x-y plane in an arbitrary direction $\alpha$. In Cartesian coordinates, we write this shift as $x \rightarrow \Tilde{x}= x + \delta r \cos\alpha$ and $y \rightarrow \Tilde{y}= y + \delta r \sin\alpha$. We consider the parameter of interest to be the magnitude of the displacement, $\delta r$, that we wish to estimate. This parameter is encoded in the mode structure of the light rather than the quantum state itself, so we use the mode parameter estimation techniques developed in \cite{Gessner:23}. This section is dedicated to reviewing the main results of this reference within the context of beam displacement estimation. 

The state of the quantum light, $\hat{\rho}$, is defined on electromagnetic field modes forming an orthonormal mode basis we denote by $\{f_k\}_k$ with an inner product $(f_j|f_k)=\delta_{jk}$. Experimentally, we are looking to detect small deviations of $\delta r$ around a given working point, which we fix to be $\delta r=0$. These lead to perturbations in the modes dependent on the parameter, resulting in a shifted mode basis $\{f_k(\delta r)\}_k$. Such changes in the parameter also induce an evolution of the quantum state that is described by a unitary beam splitter with effective Hamiltonian~\cite{Gessner:23}
\begin{equation}
    \hat{H}_{\mathrm{eff}} = i\sum_{j,k} (f_j|f'_k)\hat{a}^\dagger_j \hat{a}_k \,.
\end{equation}
In the above, we introduced the canonical creation and annihilation operators. In our context, $\hat{a}_j^\dagger$ creates a photon in the mode $f_j$, while $\hat{a}_k$ annihilates a photon from mode $f_k$. Also, we denote by $f'_k = \frac{\partial}{\partial \delta r} f_k \big|_{\delta r=0}$ the derivative mode of $f_k$. The above generator corresponds to the transformation that imprints the parameter in the initial quantum state, which itself is assumed to be parameter-independent. 

The sensitivity of the quantum state to the variations of the parameter can be quantified by the quantum Fisher information (QFI), denoted $F_Q[\hat{\rho}(\delta r)]$~\cite{Helstrom:76,BraunsteinCaves1994}. This quantity also defines a quantum limit on the estimation error of the parameter for an optimal measurement and $\nu$ repetitions via the quantum Cramér-Rao bound~\cite{Helstrom:76,Holevo:82,Paris:09,PezzeSmerzi:14,Demkowicz:15}
\begin{equation}
    (\Delta\delta r)^2 \geq \frac{1}{\nu \, F_Q[\hat{\rho}(\delta r)]}\,.
\end{equation}
Finding the ultimate theoretical precision limit is therefore equivalent to a maximisation of the quantity $F_Q[\hat{\rho}(\delta r)]$ over input states. When we specify the set of modes, $I$, that are initially populated in the quantum state, 
the QFI has been shown to take the form~\cite{Gessner:23}
\begin{align}
    F_Q[\hat{\rho}(\delta r)] \Big|&_{\delta r =0} = F_Q[\hat{\rho}(0) , \hat{H}_I] \\
    &+4 \sum_{k,l \in I} \Big[ (f'_k|f'_l) - \sum_{j\in I} (f'_k|f_j)(f_j|f'_l) \Big] \langle\hat{a}_k^\dagger\hat{a}_l\rangle_{\hat{\rho}}\,, \nonumber
\end{align}
where $\hat{H}_I$ is the effective subspace Hamiltonian corresponding to the initially populated modes $j,k\in I$. While the second term is always linear in the photon number, the term $F_Q[\hat{\rho}(0) , \hat{H}_I]$ can reach Heisenberg scaling. Therefore, it is this subspace Hamiltonian that contains the relevant quantum fluctuations that may generate the quadratic sensitivity term. Identifying the class of initial probe states and mode structures for which this is the case, and then finding the theoretically optimal initial state is exactly our interest.

\begin{figure*}[t] 
\centering 
\includegraphics[width=\textwidth]{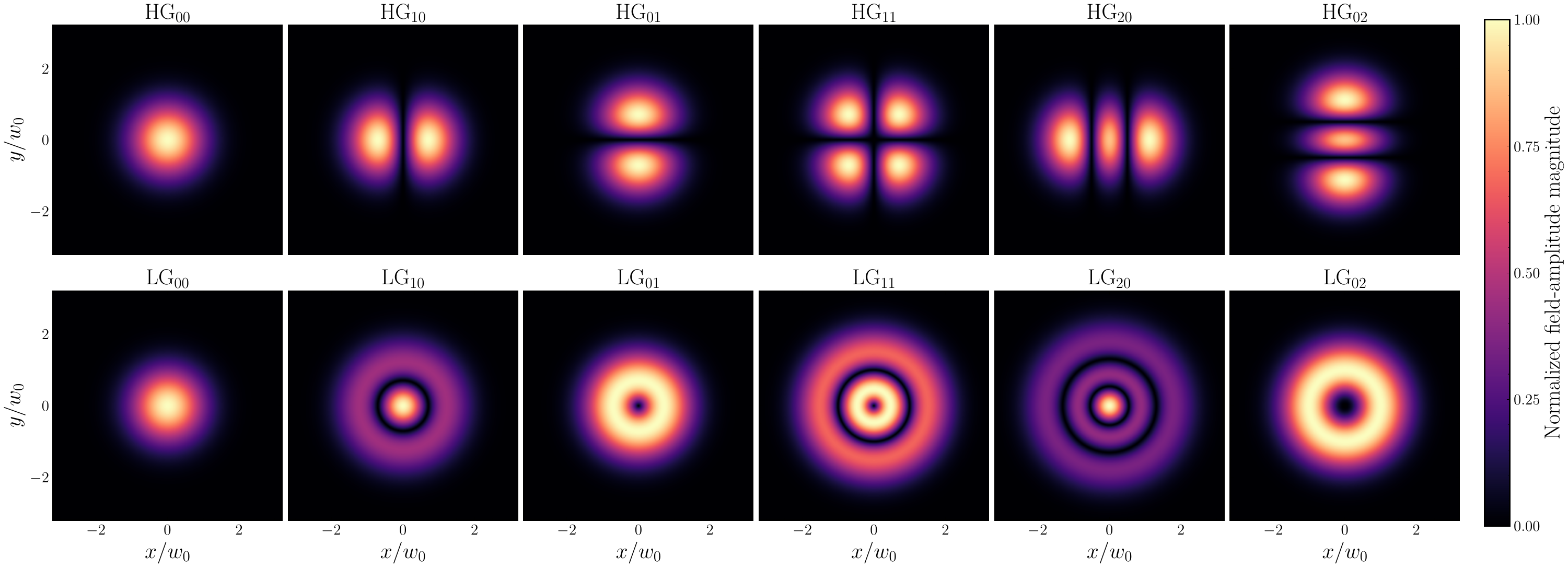} 
\caption{Illustration of selected low-order Hermite-Gauss (top row) and Laguerre-Gauss (bottom row) modes.} \label{fig: HG_LG_modes} 
\end{figure*}

The choice of mode basis is a matter of convenience for the experimenter. Two common mode bases often used in quantum optics are the Hermite-Gauss and Laguerre-Gauss modes, illustrated in Figure \ref{fig: HG_LG_modes}.

\subsection{Hermite-Gauss modes}
Hermite-Gauss (HG) modes are described by two indices $n$ and $m$ that determine the shape of the beam in the $x$ and $y$ directions respectively. They have rectangular symmetry along the propagation axis, and the first mode HG$_{00}$ is a regular Gaussian beam. The mathematical expression for an arbitrary HG mode is \cite{Siegman1986}
\begin{align}
    \mathrm{HG}_{nm}(x,y,z)
    &=
    \sqrt{\frac{2}{\pi}}
    \frac{1}{w(z)}
    \frac{1}{\sqrt{2^{m+n}m!n!}} \\ & \quad
    H_n\!\left(\frac{\sqrt{2}\,x}{w(z)}\right)
    H_m\!\left(\frac{\sqrt{2}\,y}{w(z)}\right) \nonumber\\ \nonumber
    &\quad 
    \exp\!\left[-\frac{x^2+y^2}{w^2(z)}\right]
    \exp\!\left[-i\,\frac{k(x^2+y^2)}{2R(z)}\right] \nonumber\\&\quad
    \exp\!\left[i(m+n+1)\zeta(z)\right]
    \exp(-ikz)\,, \nonumber
\end{align}
where $H_l$ is the $l$th Hermite polynomial, $w(z)$ is the beam radius, $R(z)$ is the radius of curvature, $\zeta(z)$ is the Gouy phase and $k$ is the wavenumber. We construct our orthonormal mode basis using these HG functions at the beam waist, in other words $(\mathrm{HG}_{nm}| \mathrm{HG}_{pq})=\delta_{np}\delta_{mq}$. 

In order to find the effective Hamiltonian and QFI, we make use of the HG ladder relations \cite{Siegman1986}, which for the derivative of an arbitrary HG mode $\mathrm{HG}'_{nm} = \frac{\partial}{\partial \delta r} \mathrm{HG}_{nm} \big|_{\delta r=0}$ around the reference point of $\delta r=0$ give \cite{Gessner:23, colomersaus2022}
\begin{align}
    \mathrm{HG}'_{nm}  =& \frac{\cos\alpha}{w_0} \Big(\sqrt{n} \, \mathrm{HG}_{n-1, m} - \sqrt{n+1} \, \mathrm{HG}_{n+1, m} \Big) \nonumber\\
    &+ \frac{\sin\alpha}{w_0} \Big(\sqrt{m} \, \mathrm{HG}_{n,m-1} - \sqrt{m+1} \, \mathrm{HG}_{n,m+1} \Big) \,.
\end{align}
We can then write the inner products
\begin{align}
    &\qquad(\mathrm{HG}_{nm}|\mathrm{HG}'_{pq}) \notag\\&= \frac{\cos\alpha}{w_0}
\left[\sqrt{n+1}\,\delta_{n+1,p}\delta_{mq}-\sqrt n\,\delta_{n-1,p}\delta_{mq}
\right] \nonumber \\
&\quad+\frac{\sin\alpha}{w_0}\left[\sqrt{m+1}\,\delta_{np}\delta_{m+1,q}-\sqrt m\,\delta_{np}\delta_{m-1,q}\right]. \label{eq:HG_overlap}
\end{align}
Therefore, an effective subspace Hamiltonian is nonzero only for mode pairs that differ as
\begin{align}
    &p=n\pm1 \,, \quad \text{and} \quad q=m  \quad \text{or} \label{eq: HG rules 1}\\
    &p=n \,, \quad \text{and} \quad q=m\pm1 \,. \label{eq: HG rules 2}
\end{align}
The effective Hamiltonian then has the form, as obtained in \cite{Gessner:23} and \cite{colomersaus2022},
\begin{align}\label{eq: HG general Heff}
    \hat{H}_{\mathrm{eff}} = \frac{i}{w_0}& \sum_{n,m}\Bigg( \cos\alpha \sqrt{n+1} \big( \hat{a}^\dagger_{nm} \hat{a}_{n+1,m} - \hat{a}^\dagger_{n+1,m} \hat{a}_{nm} \big) \nonumber \\
    &+ \sin\alpha \sqrt{m+1} \big( \hat{a}^\dagger_{nm} \hat{a}_{n,m+1} - \hat{a}^\dagger_{n,m+1} \hat{a}_{nm} \big) \Bigg) \,.
\end{align}
An equivalent expression including propagation-dependent complex phases and mode-dependent transmissivities was also derived in \cite{He:26}. 

It is worth stressing that the selection rules \eqref{eq: HG rules 1}-\eqref{eq: HG rules 2} refer to modes centred at the working point. Since transverse displacements form a one-parameter group, the generator \eqref{eq: HG general Heff} is independent of $\delta r$. In a mode basis recentred at any working point $\delta r_0$ it has the same coupling structure, and the QFI is unchanged. If instead the modes are referred to a fixed lab-frame basis, the populated modes at $\delta r_0 \neq 0$ spread over many HG modes (the displaced $\text{HG}_{00}$ is a coherent state of the HG ladder), and the two-mode description holds only for $|\delta r_0| \ll w_0/\sqrt{N}$. This lab-frame picture is the relevant one for the measurements of Sec. \ref{sec: Measurement schemes}, which is why we work at $\delta r_0 = 0$.

Notice that for each pair $n,m$ we have contributions to the effective Hamiltonian from and to the contiguous modes $\mathrm{HG}_{n+1,m}$ and $\mathrm{HG}_{n,m+1}$. When considering that only a finite set of modes are populated, we observe leakage to and from vacuum modes, which contributes linearly to the QFI. We also observe that populating a single mode is not sufficient to generate a non-zero effective subspace Hamiltonian. Indeed, using Eq.~\eqref{eq:HG_overlap}, it is clear that $(\mathrm{HG}_{nm}| \mathrm{HG}'_{nm})=0$, and hence $\hat{H}_I=i(\mathrm{HG}_{nm}| \mathrm{HG}'_{nm})\hat a_{nm}^\dagger\hat a_{nm}=0$. To obtain a non-zero subspace Hamiltonian, a necessary but not sufficient condition for Heisenberg scaling, at least two modes need to be populated initially. The result that a single mode is insufficient to generate quantum enhancements was first pointed out in \cite{Fabre:00}. 

Let us now restrict ourselves to initial states with only two populated modes. These modes are chosen such that the two-mode subspace Hamiltonian is non-vanishing, so if one mode is $\text{HG}_{nm}$, the other must be either $\text{HG}_{n\pm1,m}$ or $\text{HG}_{n,m\pm1}$. The choice of plus or minus is equivalent after relabeling. Notice that, depending on our choice, the subspace Hamiltonian still vanishes for certain displacement angles $\alpha$. In other words, the choice of the second mode determines an axis on which this setup is significantly less sensitive to displacements. Conversely, on the orthogonal axis the sensitivity is greatly improved. This also means that the choices $\text{HG}_{n\pm1,m}$ and $\text{HG}_{n,m\pm1}$ are equivalent under $\frac{\pi}{2}$ rotation of the coordinates. Furthermore, when the displacement direction is known, the second populated mode can always be aligned with this optimal axis, i.e. 
\begin{equation}
    \mathrm{HG}_{\mathrm{mix}}^{\mathrm{opt}}
=\frac{\sqrt{n+1}\cos\alpha\,\mathrm{HG}_{n+1,m}
+\sqrt{m+1}\sin\alpha\,\mathrm{HG}_{n,m+1}
}{\sqrt{(n+1)\cos^2\alpha+(m+1)\sin^2\alpha}} .
\end{equation}
This moves the angle dependence from the sensitivity into the populated mode.

Let us choose the initially populated modes to be $\text{HG}_{nm}$ and $\text{HG}_{n+1,m}$ and define $\hat{a}:=\hat{a}_{nm}$ and $\hat{b}:=\hat{a}_{n+1,m}$. The effective subspace Hamiltonian then has the form
\begin{equation}\label{eq: HG Heff}
    \hat{H}_I = \frac{i\sqrt{n+1}}{w_0} \cos\alpha \Big(\hat{a}^\dagger \hat{b} - \hat{b}^\dagger \hat{a} \Big) \,,
\end{equation}
and the QFI for a two-mode initial state $\hat{\rho}$ becomes
\begin{align}\label{eq: HG QFI}
    F_Q[\hat{\rho}(\delta r)] = &\:F_Q[\hat{\rho}(0), \hat{H}_I] \nonumber \\
    &+ \frac{4}{w_0^2} \Bigg( \big(n \cos^2\alpha + (2m+1) \sin^2\alpha \big) \langle\hat{a}^\dagger \hat{a}\rangle \\
    &+ \big( (n+2) \cos^2\alpha + (2m+1) \sin^2\alpha \big) \langle\hat{b}^\dagger \hat{b}\rangle \Bigg) \,. \nonumber
\end{align}

\subsection{Laguerre-Gauss modes}
While HG modes are naturally adapted to Cartesian coordinates, Laguerre-Gauss (LG) modes provide an alternative complete basis adapted to the cylindrical symmetry of paraxial propagation. Labelled by radial and azimuthal indices, the LG modes are given by \cite{Siegman1986}
\begin{equation}
\begin{aligned}
\mathrm{LG}_{nm}(r,\theta,z)
&=
\sqrt{
\frac{2n!}{\pi (n+|m|)!}
}
\frac{1}{w(z)}
\left(
\frac{\sqrt{2}r}{w(z)}
\right)^{|m|}
\\
&\quad L^{|m|}_{n}
\left(
\frac{2r^2}{w^2(z)}
\right)
\exp\left[
-\frac{r^2}{w^2(z)} -\frac{ikr^2 z}{2(z^2+z_R^2)}
\right]
\\&\quad\exp\left[
i(2n+|m|+1)\zeta(z)
\right]
\\[2mm]
&\quad 
\exp(im\theta)
\exp(-ikz)\,,
\end{aligned}
\end{equation}
with $L^{|m|}_{n}$ the generalised Laguerre polynomials. The transverse displacement transformation is expressed as
\begin{equation}
\begin{aligned}
r
&=
\sqrt{x^2+y^2}
\longrightarrow
\widetilde{r}
=
\sqrt{
\left(x+\delta r\cos\alpha\right)^2
+
\left(y+\delta r\sin\alpha\right)^2
},
\\[2mm]
\theta
&=
\arctan\left(\frac{y}{x}\right)
\longrightarrow
\widetilde{\theta}
=
\arctan\left(
\frac{y+\delta r\sin\alpha}
{x+\delta r\cos\alpha}
\right).
\end{aligned}
\end{equation}

Writing the LG modes at the beam waist as 
\begin{equation} 
\mathrm{LG}_{nm}(r,\theta) = R_n^{|m|}(r)e^{im\theta}, 
\end{equation} 
where $n\in\mathbb N_0$ and $m\in\mathbb Z$, the directional derivative can be decomposed as 
\begin{align} 
\partial_{\delta r} ={}& \frac{e^{i\alpha}}{2}e^{-i\theta} \left( \partial_r-\frac{i}{r}\partial_\theta \right) + \frac{e^{-i\alpha}}{2}e^{i\theta} \left( \partial_r+\frac{i}{r}\partial_\theta \right). 
\end{align} 
Its action on an LG mode is therefore $\mathrm{LG}_{nm}'=\frac{\partial}{\partial \delta r} \mathrm{LG}_{nm} \big|_{\delta r=0}$ which we compute as
\begin{align} 
\mathrm{LG}'_{nm} ={}& \frac{e^{i\alpha}}{2} e^{i(m-1)\theta} \left[ \frac{dR_n^{|m|}}{dr} + \frac{m}{r}R_n^{|m|} \right] \nonumber\\
&+ \frac{e^{-i\alpha}}{2} e^{i(m+1)\theta} \left[ \frac{dR_n^{|m|}}{dr} - \frac{m}{r}R_n^{|m|} \right]. \label{eq:LG_derivative_general} 
\end{align}
This expression is valid for both positive and negative azimuthal indices. In particular, the diagonal derivative overlap vanishes, since its angular dependence is proportional to $e^{\pm i\theta}$: 
\begin{equation} 
(\mathrm{LG}_{nm}|\mathrm{LG}'_{nm})=0. 
\end{equation} 
Consequently, the restricted subspace Hamiltonian vanishes if only one conventional LG basis mode is populated.

As before, we restrict ourselves to the two-mode scenario. From the above result, we know that the diagonal terms of the effective subspace Hamiltonian vanish, and we are left with 
\begin{equation}\label{eq: LG Heff}
    \hat{H}_I = i \Big( \kappa \, \hat{a}^\dagger \hat{b} - \kappa^* \, \hat{b}^\dagger \hat{a} \Big) \,,
\end{equation}
where we defined $\kappa := (\mathrm{LG}_{nm}|\mathrm{LG}'_{pq})$. Introducing the circular occupation numbers 
\begin{align} 
n_\pm &= n+\frac{|m|\pm m}{2}, & p_\pm &= p+\frac{|q|\pm q}{2}, 
\end{align} 
the coupling coefficient can be evaluated as 
\begin{align} 
\kappa = \frac{(-1)^{n+p}}{\sqrt{2}w_0}\Big[ &e^{i\alpha}\sqrt{p_+}\, \delta_{n_+,p_+-1}\delta_{n_-,p_-} \nonumber\\ 
&+ e^{-i\alpha}\sqrt{p_-}\, \delta_{n_+,p_+}\delta_{n_-,p_--1} \nonumber\\ 
&- e^{-i\alpha}\sqrt{p_++1}\, \delta_{n_+,p_++1}\delta_{n_-,p_-} \nonumber\\ 
&- e^{i\alpha}\sqrt{p_-+1}\, \delta_{n_+,p_+}\delta_{n_-,p_-+1} \Big]. 
\end{align}
The angular integral vanishes unless the azimuthal indices differ by one, while the radial overlap requires the total transverse orders to differ by one. Thus, a potentially nonzero coupling requires 
\begin{equation}
    q = m \pm 1 \quad \text{and} \quad 2p+|q|=2n+|m|\pm1 \,.
\end{equation}
These are the only mode-index selection rules for our chosen working point of $\delta r =0$.

To find the linear leakage contributions in the QFI, we first obtain the norms of the derivative modes, which can be computed as
\begin{align} 
(\mathrm{LG}'_{nm}|\mathrm{LG}'_{nm}) ={}& \int_0^{2\pi}d\theta \int_0^\infty r\,dr\, \Bigg| \frac{\sin(\alpha-\theta)}{r} \partial_\theta \mathrm{LG}_{nm}  \nonumber\\
& \qquad + \cos(\alpha-\theta) \partial_r \mathrm{LG}_{nm} \Bigg|^2 \nonumber\\ 
={}& \frac{2n+|m|+1}{w_0^2}, 
\end{align}
where we used that the radial-angular cross terms vanish and that
\begin{equation}
    \int_0^{2\pi} \sin^2(\alpha-\theta) d\theta = \int_0^{2\pi} \cos^2(\alpha-\theta) d\theta = \pi \,.
\end{equation}
The QFI can then be written
\begin{align}\label{eq: LG QFI}
    F_Q [\hat{\rho}] = F_Q[\hat{\rho},& \hat{H}_I] \nonumber \\
    + &4 \Bigg( \Big(\frac{2n + |m| +1}{w_0^2} - |\kappa|^2 \Big)\langle\hat{a}^\dagger \hat{a}\rangle \nonumber \\
    &+ \Big(\frac{2p + |q| +1}{w_0^2} - |\kappa|^2 \Big) \langle\hat{b}^\dagger \hat{b}\rangle \Bigg) \,.
\end{align}

\section{Theoretical limits} \label{sec: limits}

The above framework is powerful in revealing specific spatial mode designs that are necessary for quantum advantage in the limit of small beam displacement. It further allows us to analyse the natural next question: given a pair of initially populated modes, what probe state, $\hat{\rho}$, is optimal and what is the ultimate precision limit? We turn to this question now focusing first on the scenario when the total photon number is fixed before turning to pure Gaussian states.

To do this, we start by rewriting the effective two-mode subspace Hamiltonian and the QFI. First, note that for both mode bases they take the form
\begin{align}
    \hat{H}_I &= \mu \, \hat{a}^\dagger \hat{b} + \mu^* \, \hat{b}^\dagger \hat{a} \,, \\
    F_Q[\hat{\rho}(\delta r)] &= F_Q[\hat{\rho}(0), \hat{H}_I] + d_1 \langle\hat{a}^\dagger \hat{a}\rangle + d_2 \langle\hat{b}^\dagger \hat{b}\rangle \,.
\end{align}
Convexity of the QFI ensures that it is maximised by pure states, in which case it simplifies to 4 times the variance of the generating Hamiltonian \cite{Paris:09,PezzeSmerzi:14}. Applied to our case, the maximum QFI then has the form 
\begin{equation}
    \max_{\hat{\rho}} F_Q = \max_\psi \Big[ 4 \, \Delta^2_\psi \hat{H}_I + d_1 \langle\hat{a}^\dagger \hat{a}\rangle_\psi + d_2 \langle\hat{b}^\dagger \hat{b}\rangle_\psi \Big] \,.
\end{equation}

It is convenient to work in the basis that diagonalises the subspace Hamiltonian. To do this, we write $\mu = |\mu|e^{i\phi}$ and introduce new operators defined as
\begin{equation}
    \hat{c}_\pm := \frac{\hat{a} \pm e^{i\phi} \hat{b}}{\sqrt{2}} \,,
\end{equation}
with number operators $\hat{N}_\pm := \hat{c}_\pm^\dagger\hat{c}_\pm$, and total particle number $\hat{N}=\hat{N}_+ + \hat{N}_-$. The effective subspace Hamiltonian can then be expressed in the diagonal form 
\begin{equation}\label{eq:Heff}
    \hat{H}_I = |\mu| \big( \hat{N}_+ - \hat{N}_- \big) \,,
\end{equation}
and the eigenstates are Fock states in the $\hat{c}_\pm$ basis with eigenvalues $E_{N_\pm} = |\mu|(N_+ - N_-)$.  

We can make further simplifications using the Schwinger representation \cite{Schwinger:52,Chaturvedi06}. Writing
\begin{align}
    \hat{J}_x := \frac{\hat{c}_+^\dagger\hat{c}_- + \hat{c}_-^\dagger\hat{c}_+}{2} & \,, \quad \hat{J}_y := \frac{\hat{c}_+^\dagger\hat{c}_- - \hat{c}_-^\dagger\hat{c}_+}{2i} \,, \nonumber \\ \hat{J}_z :=& \frac{\hat{c}_+^\dagger\hat{c}_+  - \hat{c}_-^\dagger\hat{c}_-}{2} \,,
\end{align}

the effective subspace Hamiltonian becomes
\begin{equation}
    \hat{H}_I = 2|\mu| \hat{J}_z \,.
\end{equation}
Furthermore, notice that we can write 
\begin{align}
    \hat{a}^\dagger \hat{a} = \frac{\hat{N}}{2} +\hat{J}_x \,, \qquad \hat{b}^\dagger \hat{b} = \frac{\hat{N}}{2} -\hat{J}_x \,.
\end{align}
Thus, we can now express the maximal QFI in the Schwinger basis
\begin{align}\label{eq: max Fq}
    \max_{\hat{\rho}} F_Q = \max_\psi \Big[& 16 |\mu|^2 \, \Delta^2_\psi \hat{J}_z  \\ \nonumber &+ (d_1 - d_2) \big\langle \hat{J}_x \big\rangle_\psi + \frac{d_1 + d_2}{2} \big\langle \hat{N} \big\rangle_\psi \Big]  \,.
\end{align}

We may now look for optimal probe states and their sensitivities in the presence of our chosen resource constraints, and assess examples using the two-mode bases. 

\subsection{Fixed photon number} \label{sec: fixed N opt}

As the first family of states, assume that the total photon number is fixed, given by $N$. This allows us to write the QFI of a pure state $|\psi\rangle$ as
\begin{equation}
    F_Q[\hat{\rho}_\psi] =   \big\langle \hat{K} \big\rangle_{\psi} -16 |\mu|^2 \big\langle \hat{J}_z \big\rangle_{\psi}^2 + \frac{d_1 + d_2}{2} N   \,,
\end{equation}
with 
\begin{equation}
    \hat{K} := 16 |\mu|^2 \hat{J}^2_z + (d_1 - d_2) \hat{J}_x \,.
\end{equation}
For any initial state, the following inequalities hold
\begin{align}
    \big\langle \hat{K} \big\rangle_\psi &\leq \lambda_{\mathrm{max}}(\hat{K}) \,, \\
    -16 |\mu|^2 \big\langle \hat{J}_z \big\rangle^2_\psi &\leq 0 \,.
\end{align}
In the above, we have introduced the maximum eigenvalue of the operator $\hat{K}$, denoted $\lambda_{\mathrm{max}}(\hat{K})$. This eigenvalue is expected to scale quadratically with the total photon number due to $\langle \hat{J}^2_z \rangle$, providing the quantum enhancement. From these inequalities, we must have 
\begin{equation}
    F_Q[\hat{\rho}_\psi] \leq \lambda_{\mathrm{max}}(\hat{K}) + \frac{d_1 + d_2}{2} N \,.
\end{equation}
This bound turns out to be attainable. To see this, first notice that the operator $\hat{K}$ has parity symmetry in the Dicke basis, i.e. it is parity-even since both $\hat{J}_x$ and $\hat{J}_z^2$ are parity-even. Let $\ket{j,k} := \ket{j+k, j-k}_{\hat{c}_\pm}$ be the Dicke state with $j := N/2$ the total angular momentum quantum number given by half the total photon number. Then, defining the parity operator as $\hat{\Pi}\ket{j,k} = \ket{j,-k}$, it can be easily verified that $\hat{\Pi}\hat{J}_x\hat{\Pi} = \hat{J}_x$, $\hat{\Pi}\hat{J}_z\hat{\Pi} = -\hat{J}_z$ and $\hat{\Pi}\hat{J}_z^2\hat{\Pi} = \hat{J}_z^2$. Therefore, $\big[\hat{K},\hat{\Pi}\big]=0$, meaning that the eigenstates of $\hat{K}$ can be chosen to have definite parity. In other words, the eigenstate of $\hat{K}$ with maximum eigenvalue can always be chosen to be also an eigenstate of $\hat{\Pi}$. For any such parity eigenstate, the spin-$z$ operator has odd parity and thus $\langle \hat{J}_z \rangle_\psi =0$ \cite{Schwinger:52, Chaturvedi06}. The optimal states are therefore such eigenstates and can be expressed in the Dicke basis as
\begin{equation}
    \ket{\psi_{\mathrm{opt}}} = \sum^{j}_{k=-j} v_k^{(\mathrm{max})} \ket{j,k}\,,
\end{equation}
where the Schwinger spin is identified to be $j=\frac{N}{2}$ due to the fixed total photon number and $v_k^{(\mathrm{max})}$ is the $k^{\mathrm{th}}$ component of the normalised eigenvector of $\hat{K}$ corresponding to its maximum eigenvalue. Due to the parity symmetry, it can be chosen such that $v_k^{(\mathrm{max})}=\pm v_{-k}^{(\mathrm{max})}$. Therefore, in the $\hat{c}_\pm$ Fock basis, the optimal state is composed of NOON-like states 
\begin{equation}\label{eq: full opt state}
    \ket{\psi_{\mathrm{opt}}} = \sum^{j}_{k>0} u_k^\sigma \frac{\ket{j+k, j-k}+\sigma\ket{j-k,j+k}}{\sqrt{2}} + u_0^+ \ket{j,j} \,, 
\end{equation}
where $\sigma = \{+,-\}$ picks the optimal parity sector, depending on whether $\lambda_{\mathrm{max}}^+$ or $\lambda_{\mathrm{max}}^-$ is larger. This turns out to depend on the sign of $d_1 - d_2$ and whether $N$ is even or odd
\begin{equation}
    \sigma = \begin{cases}
        +1, & d_1-d_2 >0, \\ (-1)^N , & d_1-d_2 <0
    \end{cases}
\end{equation}
In the case of equality $d_1=d_2$, both sectors are equally optimal. The coefficients are related to the Dicke basis coefficients by
\begin{align}
    u_k^\sigma &= \sqrt{2} v_k^{\mathrm{max}} \,, \quad k>0 \\
    u_0^+ &= \begin{cases} 
v_0^{\mathrm{max}}, & \text{if } N \text{ is even} \,\mathrm{ and }\, \sigma=+ \\ 
0, & \text{otherwise} 
\end{cases}
\end{align}
The above optimal state was also obtained numerically in the thesis \cite{colomersaus2022}. The theoretical upper bound on the sensitivity for estimating the beam displacement distance with fixed photon number, in a given HG or LG mode pair, is then
\begin{equation}
    F_{Q}^{\mathrm{max}} [\op{\psi_{\mathrm{opt}}}{\psi_{\mathrm{opt}}} ] = \lambda_{\mathrm{max}}^\sigma + \frac{d_1 + d_2}{2} N \,.
\end{equation}

It is the first term that brings the quantum enhancement, as it carries components proportional to $N^2$, providing Heisenberg scaling. For large $N$ these contributions which originate from $\Delta^2\hat{H}_I$ will dominate. This is the variance of the $\hat{J}_z$ operator, which is known to be maximised by the NOON state~\cite{Giovannetti:06}. 

\begin{figure}[t!]
\centering
\includegraphics[width=0.99\linewidth]{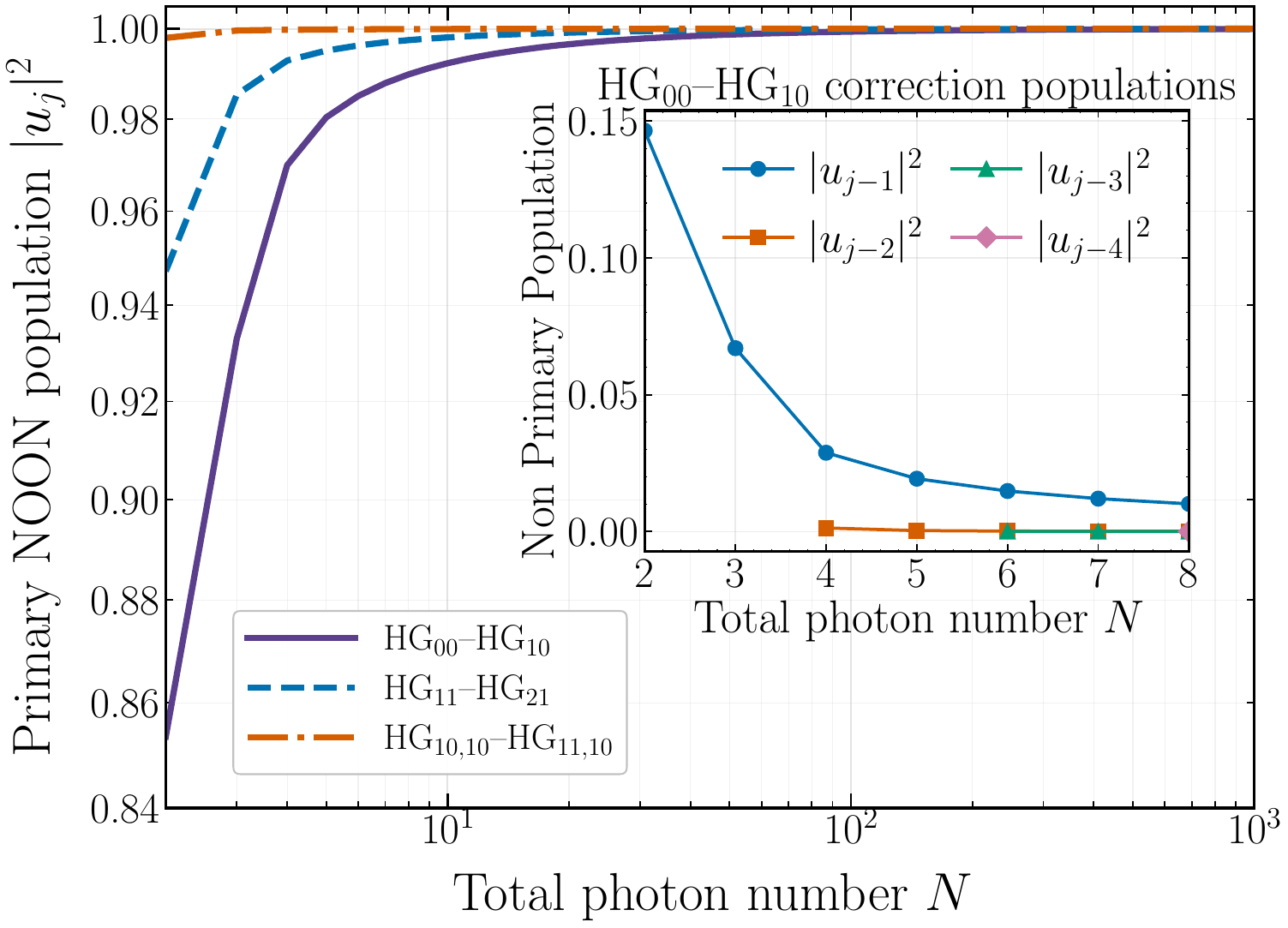}
\caption{The main panel shows the primary NOON populations for different HG mode pairs, while the inset shows the non-primary NOON-like populations for the ground and first excited HG mode pair. The populations are taken to be the absolute square of the coefficients in the optimal state expansion given in equation \eqref{eq: full opt state}. The optimal state becomes increasingly dominated by the primary NOON component as the total photon number or relevant mode index increases.}
\label{Fig: NOON Populations}
\end{figure}

In Fig.~\ref{Fig: NOON Populations}, we plot the squared absolute values of the coefficients of the NOON-like components of the expanded form of the optimal state in equation \eqref{eq: full opt state} for different total photon numbers. These are evaluated for $\cos\alpha=1$, but the optimal state populations are independent of the angle $\alpha$ (except for $\cos \alpha = 0$, for which the effective subspace Hamiltonian vanishes and there is no unique optimal population distribution). We observe that the NOON state dominates over all other NOON-like states and asymptotically becomes the optimal state. In the inset, we only considered the ground and first excited HG mode pair, but we find that higher-order mode indices lead to an even higher dominance of the NOON state and a faster asymptotic transition.

Therefore, the NOON state in the $\hat{J}_z$ basis is asymptotically optimal and constitutes a good approximation for $N >1$
\begin{equation}
    \ket{\psi_{\mathrm{NOON}}} = \frac{\ket{N,0}+e^{i\varphi}\ket{0,N}}{\sqrt{2}}\,, \quad \forall\varphi \,.
\end{equation}
In the original basis, this can be expressed as a rotated NOON state. The corresponding QFI shows Heisenberg scaling and is given by 
\begin{equation}
    F_Q [\hat{\rho}_{\mathrm{NOON}}] = 4 |\mu|^2 N^2 + \frac{d_1 + d_2}{2} N \,, \qquad N > 1 \,.
\end{equation}
For $N=1$ there is an additional term coming from $\big\langle \hat{J}_x \big\rangle_{1001}= \frac12 \cos\varphi$, which is 0 for all higher $N$. In the single-photon special case, however, we have 
\begin{equation}
    F_Q[\hat{\rho}_{1001}] = 4 |\mu|^2  + \frac{d_1 + d_2}{2} + \frac{d_1 - d_2}{2} \cos\varphi\,.
\end{equation}
This additional term still vanishes for a purely imaginary phase, and conversely, the single-photon NOON QFI is optimal for a real phase with $\cos \varphi_{\mathrm{opt}}^{N=1} = \mathrm{sgn}(d_1 - d_2)$.   

 \begin{figure}[t!]
\centering
\includegraphics[width=0.99\linewidth]{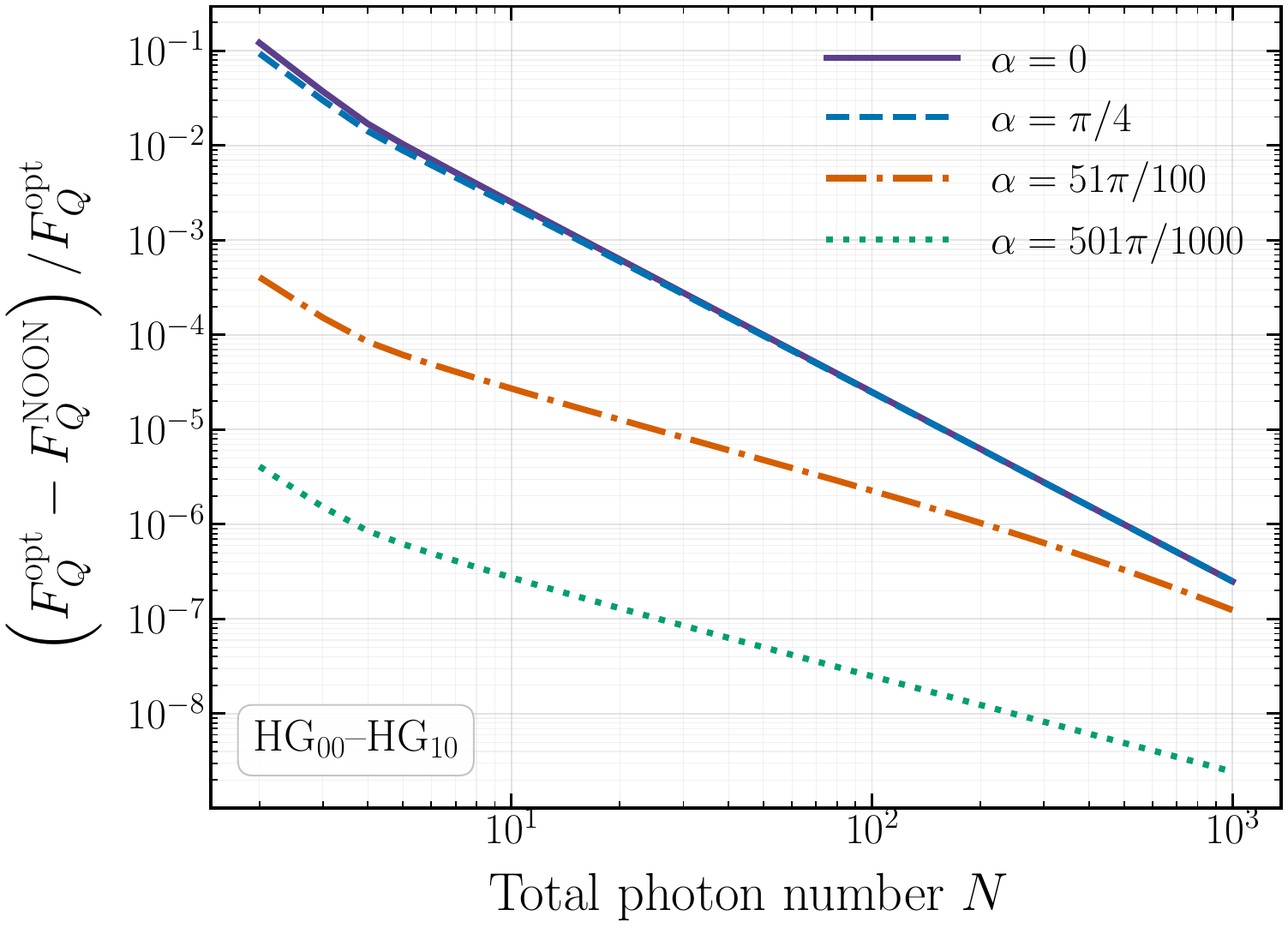}
\caption{The relative difference between the true maximum sensitivity and the asymptotically optimal NOON state sensitivity against photon number starting at $N=2$ for different displacement directions using the HG mode pair. For $N=1$ the NOON state is optimal and the relative difference collapses.}
\label{Fig: Opt-NOON QFI}
\end{figure}

Figure \ref{Fig: Opt-NOON QFI} shows the relative difference between the ultimate maximum QFI using the NOON-like expanded state (evaluated numerically) and the QFI corresponding to the NOON state, the leading-order term of the expansion. The plot is for $N>1$; for the single-photon case, the relative difference is 0 and the NOON state with real phase is optimal. For this comparison, the mode pair $\text{HG}_{00}$ and $\text{HG}_{10}$ was used, and the graph shows plots for different displacement directions characterised by the displacement angle $\alpha$. The angle $\alpha = \pi/4$ corresponds exactly to the mode pair $\text{LG}_0^0$ and $\text{LG}_0^1$. In all cases, the difference decays rapidly as the number of photons increases, and the NOON state asymptotically converges to the optimal state in relative terms. As $\alpha$ increases $|\mu|$ decreases, leading to a significant drop in the absolute sensitivities and a slight difference in relative sensitivities for small $N$. At the angle $\alpha = \pi/2$, the effective subspace Hamiltonian vanishes and the QFI has only linear terms. At that point, the difference collapses and all $N$-photon states are equally optimal.

\subsection{Pure Gaussian states}\label{sec: Gaus opt}

Let us now consider two-mode Gaussian probes with a fixed mean photon number, $\Bar{N}$. Moreover, we restrict ourselves to the pure Gaussian family of states that are also centred, i.e. have vanishing first quadrature moments. These states have been shown to be optimal for our two-mode, passive linear unitary \cite{Matsubara_2019}. Therefore, they provide an indicative benchmark for the optimal fixed-$N$ case. Again, it is convenient to define the orthogonal operators
\begin{align}
    \hat{g}_+ := u_+ \hat{c}_+ + u_- \hat{c}_- \,, \quad \hat{g}_- := -u_-^* \hat{c}_+ + u_+^* \hat{c}_- \,,
\end{align}
with
\begin{equation}
    |u_+|^2 + |u_-|^2 =1 \,.
\end{equation}
This way, by the Bloch-Messiah decomposition \cite{Braunstein:05}, we can express any centred pure two-mode Gaussian state as a pair of single-mode squeezed vacuum states
\begin{equation}
    \ket{\psi^{(\mathrm{G})}} = \hat{S}_{g_+} (\zeta_+) \hat{S}_{g_-} (\zeta_-) \ket{0} \,.
\end{equation}

At fixed mean photon number, the leading-order term of the QFI, given by the variance of the effective subspace Hamiltonian~(\eqref{eq:Heff}), is maximised by the squeezed vacuum states in which only the $\hat{g}_+$ mode is populated~\cite{Monras_2006, Fadel_2025, Matsubara_2019}. This state is therefore asymptotically optimal for large $\Bar{N}$, at least within the two-mode family of Gaussian states. We now evaluate the QFI for this squeezed Gaussian state. We parameterise the coefficients as
\begin{equation}
    u_+ = \cos \frac{\beta}{2} \,, \quad u_- = e^{i\gamma} \sin \frac{\beta}{2} \,.
\end{equation}
Note that any overall phase can be absorbed into the squeezing phase, which rotates the state in quadrature space and is consequently absorbed into the optimal measurement. Thus, only the relative phase, $\gamma$, needs to be considered for the QFI. So, for the squeezed state 
\begin{equation}
    \ket{\psi_{\mathrm{sq}}} = \hat{S}_{g_+} (\zeta) \ket{0} \,,
\end{equation}
with $\zeta = r_g e^{i\vartheta_g}$ and $\Bar{N} = \sinh^2 r_g$, we can compute the terms of Eq. \eqref{eq: max Fq} and find
\begin{align}
    \Delta^2 \hat{J}_z &= \frac{\Bar{N}}{4} \big( 1 + (2\Bar{N} + 1)\cos^2\beta \big) \,, \\
    \langle \hat{J}_x \rangle &= \frac12 \Bar{N} \sin\beta \cos\gamma\,, \\
    \langle \hat{N} \rangle &= \Bar{N} \,.
\end{align}
The term $(d_1-d_2)\langle \hat{J}_x \rangle$ is maximal for phase $\gamma=0$ or $\pi$, and it has no effect on any other terms of the QFI. This leads to the conclusion that the optimal coefficients are always real, and the choice of $e^{i\gamma}=\pm 1$ can be captured in $\beta$. The maximum QFI for these Gaussian states is therefore 
\begin{align}
    F_{Q}^{\mathrm{max}} [\hat{\rho}_{\mathrm{sq}} ] =\:& 8|\mu|^2 \Bar{N} (\Bar{N}+1) - 4|\mu|^2 \Bar{N} (2\Bar{N}+1)\sin^2\beta \nonumber \\
    &+ \Big( \frac{d_1 + d_2}{2} + \frac{d_1 - d_2}{2} \sin\beta \Big)\Bar{N}
\end{align}
This is a concave function quadratic in $\sin\beta$, which can be solved to find the optimal mixture of modes, $\beta_{\mathrm{opt}}$. The maximum is easily found to be at 
\begin{equation}
    \sin\beta_{\mathrm{opt}} = \begin{cases} 
    \frac{d_1-d_2}{16|\mu|^2 (2\Bar{N}+1)} \,, & \Big| \frac{d_1-d_2}{16|\mu|^2 (2\Bar{N}+1)} \Big| \leq1, \\
    \mathrm{sgn}(d_1-d_2)\,, & \text{otherwise}.
    \end{cases}
\end{equation}
Inside the interval, the maximum QFI becomes
\begin{align}\label{eq: max Gaussian Fq}
    F_{Q}^{\mathrm{max}} [\hat{\rho}_{\mathrm{sq}} ] =\:& 8|\mu|^2 \Bar{N} (\Bar{N}+1) \nonumber \\ &+ \Big( \frac{d_1 + d_2}{2} + \frac{(d_1 - d_2)^2}{64 |\mu|^2 (2\Bar{N}+1)} \Big)\Bar{N} \,.
\end{align}

We can also see that for large enough average photon numbers the asymptotically optimal Gaussian state is well approximated by a single-mode squeezed state in just the $\hat{c}_+$ mode (i.e. $u_- \approx 0$). Therefore, for the two-mode Gaussian states, the asymptotically optimal state is a single-mode squeezed state in the $\hat{c}_+$ mode.

\subsection{Examples: Ground and first excited modes }\label{sec: examples}

As an example, let us consider the case when the two initially populated modes are the ground and first excited modes. That is, in the HG basis we populate the pair $\text{HG}_{00}$ and $\text{HG}_{10}$, and in the LG basis $\text{LG}_{00}$ and $\text{LG}_{01}$. For the HG modes, we can immediately read off the relevant coefficients from Eqs.~\eqref{eq: HG Heff} and \eqref{eq: HG QFI} to be
\begin{align}\label{eq: mu,d1,d2}
    \mu^{\mathrm{(HG)}} &= \frac{\cos\alpha}{w_0}i \nonumber \\ d_1^{\mathrm{(HG)}} &= \frac{4 \sin^2\alpha}{w^2_0} \,, \qquad d_2^{\mathrm{(HG)}} = \frac{4 (1+\cos^2\alpha)}{w^2_0} \,.
\end{align}
For the LG modes, we must first evaluate the integral for $\kappa$, which gives 
\begin{equation}
    \kappa = \frac{(-1)^{n+p} }{\sqrt{2}w_0 }e^{i\alpha} \,.
\end{equation}
Thus, from Eqs.~\eqref{eq: LG Heff} and \eqref{eq: LG QFI} we obtain 
\begin{align}
    \mu^{\mathrm{(LG)}} &= \frac{(-1)^{n+p}}{\sqrt{2}w_0 }ie^{i\alpha} \,, \nonumber \\ d_1^{\mathrm{(LG)}} &= \frac{2}{w^2_0} \,, \qquad d_2^{\mathrm{(LG)}} = \frac{6}{w^2_0} \,.
\end{align}
Note that $|\mu^{\mathrm{(LG)}}|^2, d_1^{\mathrm{(LG)}}, d_2^{\mathrm{(LG)}}$ are all equivalent to their corresponding HG expression for $\alpha = \pi/4$. Therefore, we proceed considering only the HG modes for this example, of which the LG modes are a special case. We can then write
\begin{align}
    \hat{K}^{\mathrm{(HG)}} &= \frac{8 \cos^2\alpha}{w_0^2} \Big(2 \hat{J}_z^2 -\hat{J}_x \Big) \,.
\end{align}

For the squeezed vacuum state, we see that 
\begin{equation}
    \Big| \frac{d_1-d_2}{16|\mu|^2 (2\Bar{N}+1)} \Big| = \frac{1}{2(2\Bar{N}+1)} \leq1 
\end{equation}
is satisfied for all $\Bar{N}$. In fact, the optimal mixture parameter $\beta$, for both the HG and LG mode pairs, is given by 
\begin{equation}
    \beta_{\mathrm{opt}} = -\sin^{-1} \Bigg( \frac{1}{2(2\Bar{N}+1)} \Bigg) \,.
\end{equation}
The maximum QFI for the fixed photon number and Gaussian scenarios become
\begin{align}
    &F_{Q}^{\mathrm{HG}} \big[\op{\psi_{\mathrm{opt}}}{\psi_{\mathrm{opt}}} \big] = \frac{8 \cos^2\alpha}{w_0^2}\lambda_{\mathrm{max}} \Big(2 \hat{J}_z^2 -\hat{J}_x \Big)  + \frac{4}{w_0^2} N \,, \\
    &F_{Q}^{\mathrm{HG}} [\hat{\rho}_{\mathrm{sq}} ] = \frac{8\cos^2\alpha}{w_0^2} \Bar{N} (\Bar{N}+1) + \Big( \frac{4}{w_0^2} + \frac{\cos^2\alpha}{ w_0^2 (2\Bar{N}+1)} \Big)\Bar{N} \,. \label{eq: Fq max HG sq}
\end{align}
The corresponding results for the LG modes can be directly taken from these expressions by simply taking $\alpha=\pi/4$. The maximum eigenvalue and eigenvector in the fixed-$N$ QFI depend on the particle number, but for large photon numbers the optimal state is well approximated by the NOON state and the leading order term is $\lambda_{\mathrm{max}}^\sigma \approx N^2/2$. 

Comparing the two-mode designs, we note that the HG sensitivity is sinusoidally oscillating depending on the displacement angle. Moreover, the centre of oscillation is given by the LG sensitivity, which is itself independent of the displacement direction. Therefore, the two modes perform identically on average, with the HG modes yielding better sensitivities in the best-case scenario and worse, giving classical scaling, in the worst-case scenario. If the displacement direction is known, however, the HG modes can be engineered to consistently yield improved sensitivities over the LG modes, i.e. populating $\text{HG}_{\mathrm{mix}}$ as previously mentioned.

\begin{figure}[t!]
\centering
\includegraphics[width=0.99\linewidth]{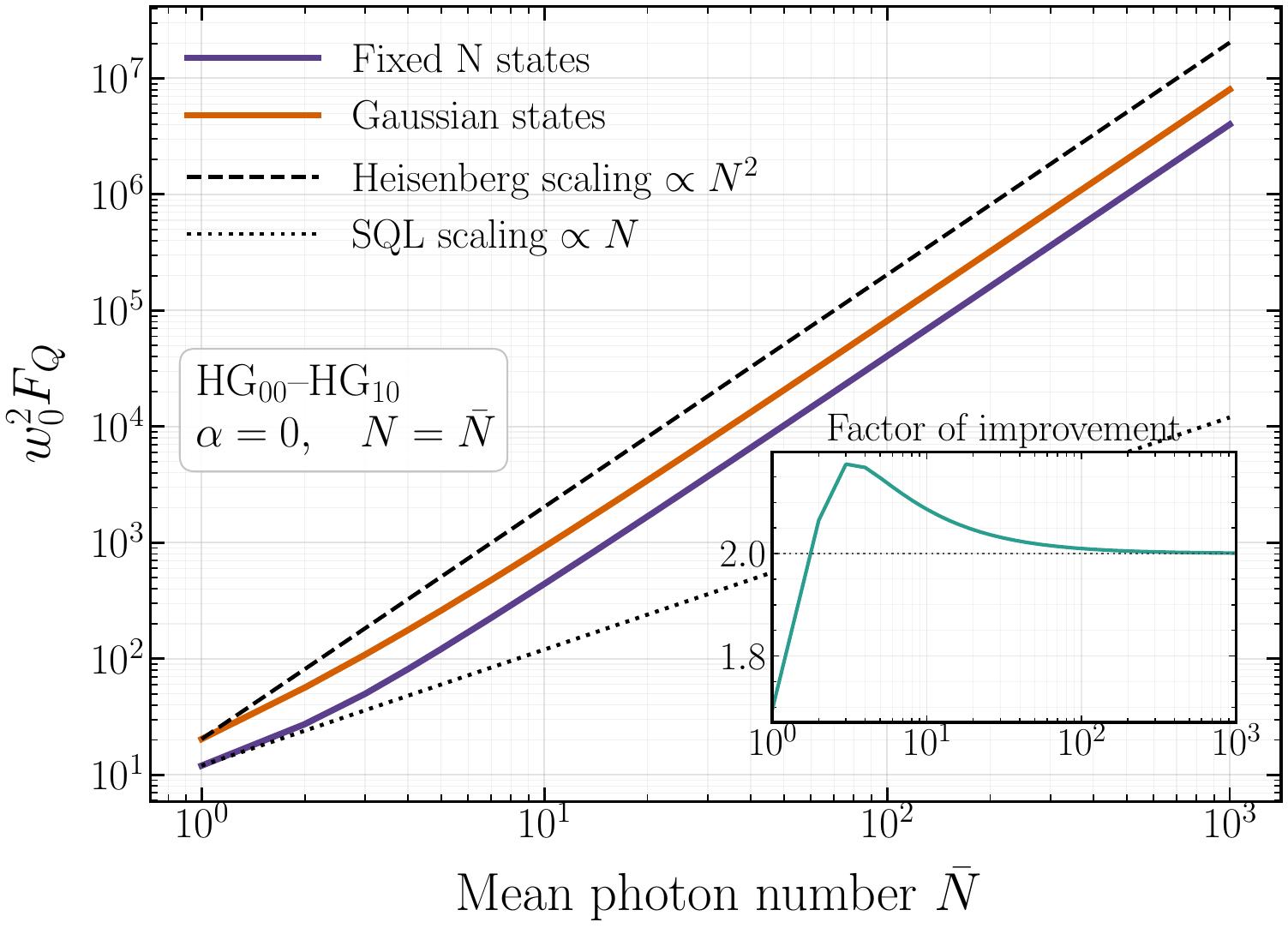}
\caption{The maximum sensitivity attainable by fixed-$N$ states using the HG mode pair with the optimal displacement direction compared to the sensitivity of the asymptotically optimal centred pure Gaussian state, taken to be the squeezed vacuum state in the $\hat{g}_+$ mode with $\beta=\beta_{\mathrm{opt}}$. The inset graph shows the factor of improvement of the Gaussian QFI over the fixed-$N$ QFI.}
\label{Fig: Gaussian vs Fixed N}
\end{figure}

We can also compare the sensitivity achievable using the asymptotically optimal Gaussian state and the optimal fixed photon number states. Figure \ref{Fig: Gaussian vs Fixed N} shows that squeezed states have the potential to outperform fixed-$N$ states even for the optimal displacement direction for the HG mode pairs. This is because, for Gaussian states, only the average photon number is fixed. Therefore, they exhibit unbounded photon number support and fluctuations, whereas the fixed-$N$ states have a fixed photon budget. The factor of improvement of the optimal Gaussian QFI over the fixed-$N$ QFI is approaching 2 for large average photon numbers. For small $N$ larger than 1, where the quadratic contributions are not as dominant, the Gaussian QFI gives an improvement slightly better than just doubling the fixed-$N$ QFI. 

\begin{figure}[t!]
\centering
\includegraphics[width=0.99\linewidth]{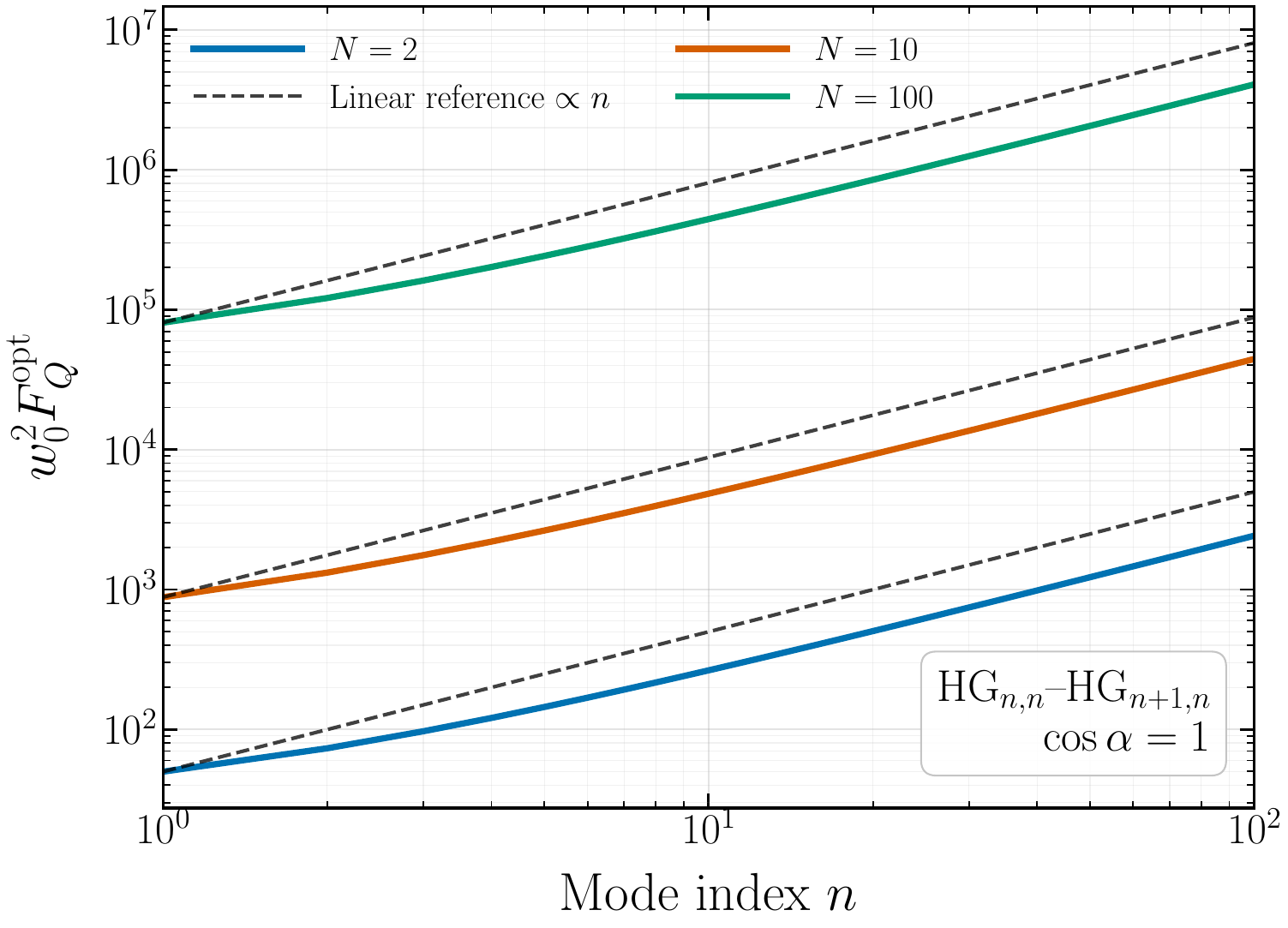}
\caption{The optimal fixed-$N$ QFI for higher-order mode pairs against the order index for different photon numbers. The angle is fixed to $\cos\alpha=1$, so $m$ becomes irrelevant and is set to $m=n$. For each QFI, a linear reference line is shown demonstrating that the scaling is asymptotically linear for high mode orders for all $N$. }
\label{Fig: Higher mode orders}
\end{figure}

Looking beyond the lowest possible mode indices, we note that $|\mu|^2$, $d_1$ and $d_2$ are all linear in the mode indices $n$ and $m$. In the special case of $\cos\alpha=1$, only the index $n$ is relevant and the choice of index $m$ does not affect the sensitivity. Going to higher-order modes therefore improves the QFI, but the scaling is at most linear in the mode index. We illustrate this in Fig.~\ref{Fig: Higher mode orders}.

It must be noted that all the results presented here assume a perfect transmission from the source to the detector. In most realistic setups, noise originating from photon losses to the environment must be taken into account. This can be modelled as a beam splitter mixing the original modes with environmental modes. Consequently, these additional modes must be traced out, leading to mixed states at the detector due to the loss of information, and thus drastically reduced noisy QFI. A full analysis is outside the scope of this paper, but for the asymptotically optimal squeezed vacuum state it has been carried out in Ref.~\cite{He:26}. 

Beyond Gaussian states, if only the average photon number is bounded, there is no fundamental limit to how large the QFI may get~\cite{Shapiro_1989,Rivas_2012,Pezze_2013,Pasquale_2015,JiaJie_25}. Exotic non-Gaussian states could be considered, but their applicability is often difficult to justify. Nonetheless, there is growing interest in superpositions of Gaussian states in the literature, which could in principle provide further quantum enhancement \cite{Caron:26,braccini2026}.

\section{Measurement schemes with the method of moments} \label{sec: Measurement schemes}

Having established the theoretical limits, we may ask how close we can feasibly get to those limits in practice. For this, we now propose concrete measurement schemes that can achieve Heisenberg scaling in their sensitivity with technologies currently available. The method of moments is a particularly practical and simple strategy for building the estimator for $\delta r$ \cite{Gessner:19,Sorelli:21a,Sorelli:21,Fadel_2025}. 

The moment-based sensitivity for state $\hat{\rho}(\delta r)$ and measurement $\hat{M}$ is given by
\begin{equation}
    \chi^{-2}[\hat{\rho}(\delta r), \hat{M}] = \frac{\big|\partial_{\delta r}\langle\hat{M}\rangle_{\hat{\rho}(\delta r)}\big|^2}{\Delta^2 \hat{M}_{\hat{\rho}(\delta r)}} \,.
\end{equation} 
The estimator is constructed for $\nu$ repeated measurements via the comparison between the sample average of the measurements of $\hat{M}$ and a previously established calibration curve, $\langle\hat{M}\rangle_{\hat{\rho}(\delta r)}$, leading to the estimator variance
\begin{equation}
    (\Delta\delta r)^2 = \frac{1}{\nu}\chi^{2}[\hat{\rho}(\delta r), \hat{M}] \,.
\end{equation}
In the asymptotic limit of many repetitions $\nu \rightarrow \infty$ this estimator can be shown to be locally optimal, recovering the QFI if the observable $\hat{M}$ is optimally chosen~\cite{Gessner:19}. For any other, possibly non-optimal observable, the moment-based sensitivity is bounded from above by the QFI~\cite{PezzeSmerziPRL2009,Gessner:19}
\begin{equation}
    \chi^{-2}[\hat{\rho}(\delta r), \hat{M}] \leq F_Q[\hat{\rho}(\delta r)] \,.
\end{equation}

Next, we calculate the moment-based sensitivities for two fixed photon number states and demonstrate that a parity measurement saturates the Heisenberg-scaling term of their corresponding QFI. In addition, we also calculate the moment-based sensitivity for a homodyne measurement on the squeezed vacuum state.

\subsection{Parity measurement on the NOON state}

Suppose we take as our initial state the asymptotically optimal multiphoton NOON state in the rotated $\hat{c}_\pm$ basis
\begin{equation}
    \ket{\psi_{\mathrm{n}}} = \frac{\ket{N,0}_{\hat{c}_\pm}+\ket{0,N}_{\hat{c}_\pm}}{\sqrt{2}}\,, \quad N>1 \,.
\end{equation}
The evolved state is then 
\begin{align}
    \ket{\psi_{\mathrm{n}}(\delta r)} = e^{-i \delta r \hat{H}_{\mathrm{eff}}} \ket{\psi_{\mathrm{n}}} \,.
\end{align}
We introduce the parity operators on the original mode bases, $\hat{P}_{a,b}$, defined by their effect on an arbitrary Fock state
\begin{equation}
    \hat{P}_{a,b} \ket{n,m}_{a,b} = (-1)^{n,m} \ket{n,m}_{a,b} \,.
\end{equation}
The effect of $\hat{P}_a$ on the operators $\hat{c}_\pm$ is such that $\hat{P}_a \hat{c}_\pm^\dagger \hat{P}_a = -\hat{c}_\mp^\dagger$, and similarly, $\hat{P}_b \hat{c}_\pm^\dagger \hat{P}_b = \hat{c}_\mp^\dagger$, giving the eigenvalues
\begin{align}
    \hat{P}_{a,b} \ket{\psi_{\mathrm{n}}} &= p_{a,b} \ket{\psi_{\mathrm{n}}}  \,, \nonumber \\
    p_a = (-1)^N \,,& \qquad p_b = 1 \,.
\end{align}

To get the moment-based sensitivity, we must first evaluate the expectation values. We use the Baker–Campbell–Hausdorff formula to make the expansion
\begin{align}
    \langle \hat{P} \rangle_{\delta r} &= \langle e^{i \delta r \hat{H}_{\mathrm{eff}}}\hat{P}e^{-i \delta r \hat{H}_{\mathrm{eff}}} \rangle = \langle\hat{P}\rangle + i \delta r \big\langle [\hat{H}_{\mathrm{eff}},\hat{P}]\big\rangle \nonumber \\
    & \qquad \qquad - \frac{\delta r ^2}{2} \big\langle [\hat{H}_{\mathrm{eff}},[\hat{H}_{\mathrm{eff}},\hat{P}]]\big\rangle + \mathcal{O}(\delta r^3) \nonumber \\
    & = p - \delta r^2 \, p\big(\langle \hat{H}_{\mathrm{eff}}^2 \rangle - p\langle \hat{H}_{\mathrm{eff}} \hat{P} \hat{H}_{\mathrm{eff}} \rangle \big) + \mathcal{O}(\delta r^3) \,.
\end{align}
In \cite{Gessner:23}, it has already been derived that
\begin{equation}
    \langle \hat{H}_{\mathrm{eff}}^2 \rangle = \langle \hat{H}_I^2 \rangle + \frac{d_1}{4} \langle\hat{a}^\dagger \hat{a}\rangle + \frac{d_2}{4} \langle\hat{b}^\dagger \hat{b}\rangle \,.
\end{equation}
It is then easy to see, using $p\langle \hat{H}_{\mathrm{eff}} \hat{P} \hat{H}_{\mathrm{eff}} \rangle = \langle \hat{P}\hat{H}_{\mathrm{eff}} \hat{P} \hat{H}_{\mathrm{eff}} \rangle$, that
\begin{align}
    p_a\langle \hat{H}_{\mathrm{eff}} \hat{P}_a \hat{H}_{\mathrm{eff}} \rangle =& -\langle \hat{H}_I^2 \rangle - \frac{d_1}{4}  \langle\hat{a}^\dagger \hat{a}\rangle + \frac{d_2}{4}  \langle\hat{b}^\dagger \hat{b}\rangle \nonumber \,, \\
    p_b\langle \hat{H}_{\mathrm{eff}} \hat{P}_b \hat{H}_{\mathrm{eff}} \rangle =& -\langle \hat{H}_I^2 \rangle + \frac{d_1}{4}  \langle\hat{a}^\dagger \hat{a}\rangle - \frac{d_2}{4} \langle\hat{b}^\dagger \hat{b}\rangle \,.
\end{align}
We know from the previous section that for the rotated NOON state $\langle \hat{H}_I^2 \rangle = |\mu|^2 N^2$ and $\langle\hat{a}^\dagger \hat{a}\rangle = \langle\hat{b}^\dagger \hat{b}\rangle = N/2$. The expectation values are therefore
\begin{align}
    \langle \hat{P}_a \rangle_{\delta r} &= (-1)^N \Big(1 - 2 \big(|\mu|^2 N^2 + \frac{d_1}{8}N \big)  \delta r^2 \Big)  + \mathcal{O}(\delta r^3) \,, \nonumber \\
    \langle \hat{P}_b \rangle_{\delta r} &=  1 - 2 \big(|\mu|^2 N^2 + \frac{d_2}{8}N\big)  \delta r^2   + \mathcal{O}(\delta r^3) \,.
\end{align}

Calculating the expectation derivatives and the variances is then straightforward, noting that $\hat{P}_a^2 = \hat{P}_b^2 = \hat{\mathbb{I}}$. The moment-based sensitivities can be evaluated in the limit as $\delta r \rightarrow 0$ using the expanded expressions above, giving
\begin{align}
    \lim_{\delta r \rightarrow 0}\chi^{-2}[\hat{\rho}_{\psi_{\mathrm{n}}}(\delta r), \hat{P}_a] =& 4 |\mu|^2 N^2 + \frac{d_1}{2} N \,, \nonumber \\
    \lim_{\delta r \rightarrow 0}\chi^{-2}[\hat{\rho}_{\psi_{\mathrm{n}}}(\delta r), \hat{P}_b] =& 4 |\mu|^2 N^2 + \frac{d_2}{2} N \,.
\end{align}

Notice that when the optimal displacement angle is such that $\sin\alpha=0$, $d_1$ vanishes for all HG mode pairs with order $\text{HG}_{0m} - \text{HG}_{1m}$. In this special case, all the linear contributions are carried by the $b$ modes, and the moment-based sensitivity with a parity measurement in these modes, $\hat{P}_b$, locally saturates the QFI. In all other scenarios, using the method of moments with a parity measurement in one of the two original modes captures only part of the linear sensitivity. Nonetheless, the quadratic term of the QFI is obtained for all mode pairs and displacement angles; therefore, the moment-based sensitivity is still approaching the QFI for large photon numbers.  

One might wonder if a measurement observable could be constructed using both parity observables that locally saturate the full QFI for all mode pairs and displacement directions in the small displacement limit. The answer is yes. The trick is to find an observable that can distinguish between the original parity sector and the evolved parity sector, including when both parities are flipped leaving their product unchanged. The observable that can do this is
\begin{equation}
    \hat{P}_{\mathrm{joint}} = \frac14 \big( \hat{\mathbb{I}} + p_a \hat{P}_a \big) \big( \hat{\mathbb{I}} + p_b \hat{P}_b \big) \,.
\end{equation}
It has eigenvalue 1 only when both modes retain their original state parity and 0 if either or both parities flip. Physically, this corresponds to measuring the parity of each mode and then recording whether both outcomes agree with their expected initial values. Importantly, the moment-based sensitivity for this observable locally saturates the full QFI
\begin{align}
    \lim_{\delta r \rightarrow 0}\chi^{-2}[\hat{\rho}_{\psi_{\mathrm{n}}}(\delta r), \hat{P}_{\mathrm{joint}}] = 4 |\mu|^2 N^2 + \frac{d_1 + d_2}{2} N \,.
\end{align}

\subsection{Parity measurement on the twin-Fock state}

Preparing the rotated NOON state in the lab is highly challenging. Let us now discuss a potentially more practical probe also exhibiting Heisenberg scaling. For the initial state, we consider the balanced twin-Fock state in the original, unrotated mode basis with $N$ even
\begin{equation}
    \ket{\psi_{\mathrm{tf}}} = \bigg|{\frac{N}{2}, \frac{N}{2}} \bigg\rangle_{a,b} \,.
\end{equation}
For this initial state, both parity operators have the same eigenvalue, $p_a=p_b$
\begin{equation}
    \hat{P}_a \ket{\psi_{\mathrm{tf}}} = \hat{P}_b \ket{\psi_{\mathrm{tf}}} = (-1)^{N/2} \ket{\psi_{\mathrm{tf}}} \,.
\end{equation}
This is not to say that measuring the two separate parities is equivalent. As for the NOON state, each is sensitive only to changes in its own mode's parity sector, leading to different sensitivities based on the difference between $d_1$ and $d_2$. In fact, we find that measuring $\hat{P}_{\mathrm{joint}}$ once again locally saturates the QFI. Using the same steps as in the previous subsection, we obtain the moment-based sensitivity in the small displacement limit 
\begin{align}
    \lim_{\delta r \rightarrow 0}\chi^{-2}[\hat{\rho}_{\psi_{\mathrm{tf}}}(\delta r), \hat{P}_{\mathrm{joint}}] = 4 \langle \hat{H}_I^2 \rangle + \frac{d_1 + d_2}{2} N \,.
\end{align}
It is easily shown that $\langle \hat{H}_I^2 \rangle = \frac12 |\mu|^2 N(N+2)$ and that the corresponding QFI is 
\begin{equation}
    F_Q [\hat{\rho}_\mathrm{tf}] = 2 |\mu|^2 N(N+2) + \frac{d_1 + d_2}{2} N \,,
\end{equation}
thus, this strategy also locally saturates the full QFI. Compared to the ultimate upper bound with the near-optimal rotated NOON state, we see that the Heisenberg term is a factor of $1/2$ smaller, with an additional linear term. Moreover, the difference between their sensitivities scales with photon number as $2|\mu|^2 N(N-2)$, and when $N=2$, the sensitivities overlap. This is because $\ket{1,1}_{a,b}$ is exactly the NOON state in the rotated $\hat{c}_\pm$ basis. Therefore, this measurement scheme is a viable alternative strategy to achieve Heisenberg scaling in the estimation of beam displacement distance in the lossless, two-mode case.

\subsection{Homodyne measurement on a squeezed vacuum state}

We now move beyond the states with fixed photon numbers and propose a measurement scheme with a fixed average photon number, $\Bar{N}$, instead. Consider the initial state to be the previously motivated single-mode squeezed vacuum state in the $\hat{c}_+$ mode, with the $\hat{c}_-$ mode unpopulated
\begin{equation}
    \ket{\psi_{\mathrm{ss}}} = \hat{S}_+ (\xi) \ket{0} \,,
\end{equation}
with the squeezing operator $\hat{S}_+$, parameter $r_s$, and angle $\vartheta$, defined as
\begin{equation}
    \hat{S}_+ (\xi)= e^{\frac12 (\xi^* \hat{c}_+^2 - \xi \hat{c}_+^{\dagger2}) } \,, \qquad \xi= r_s e^{i \vartheta} \,.
\end{equation}
The fixed mean photon number is given by 
\begin{equation}
    \Bar{N} = \langle \hat{N}_+ \rangle = \sinh^2r_s \,.
\end{equation}
We write the evolved state as
\begin{equation}
    \ket{\psi_{\mathrm{ss}}(\delta r)} = e^{-i \delta r \hat{H}_{\mathrm{eff}}} \hat{S}_+ (\xi) \ket{0} \,.
\end{equation}

For the measurement, we choose homodyne detection directly on the $\hat{c}_+$ mode with the measured quadrature 
\begin{equation}
    \hat{X}_\theta = \frac{e^{-i\theta}\hat{c}_+ + e^{i\theta}\hat{c}_+^\dagger}{\sqrt{2}} \,.
\end{equation}
However, the first quadrature moment carries no information about the parameter, as its expectation value is $0$. Thus, we must use the second moment, $\hat{X}_\theta^2$, as our measurement observable. At our working point of $\delta r =0$, the moment-based sensitivity depends on the first derivative of its expectation value and the zeroth-order variance. The derivative of the variance at the working point is given by
\begin{equation}
    \partial_{\delta r} \langle \hat{X}_\theta^2 \rangle_{\delta r =0} = i \big \langle [ \hat{H}_{\mathrm{eff}}, \hat{X}_\theta^2 ] \big\rangle_{\delta r =0} \,.
\end{equation}
However, we find that for the squeezed vacuum state it is only $\hat{H}_I$ that contributes at the zeroth order and the remaining contributions of $\hat{H}_{\mathrm{eff}}$ first modify the quadrature statistics only at second-order, $\mathcal{O}(\delta r^2)$. Therefore, we have 
\begin{equation}
    \big \langle [ \hat{H}_{\mathrm{eff}}, \hat{X}_\theta^2 ] \big\rangle_{\delta r =0} = \big \langle [ \hat{H}_I, \hat{X}_\theta^2 ] \big\rangle_{\delta r =0} \,.
\end{equation}
Since the zeroth-order variance is independent of the evolution, we see that only the effective subspace Hamiltonian is relevant for the moment-based sensitivity at $\delta r=0 $ for this initial state.

The effective subspace Hamiltonian becomes especially simple when the $\hat{c}_-$ mode remains in the vacuum 
\begin{equation}
    \hat{H}_I = |\mu| \hat{N}_+ \,.
\end{equation}
The evolution induced by this Hamiltonian transforms the mode operator as
\begin{equation}
    e^{i |\mu| \delta r \hat{N}_+} \hat{c}_+ e^{-i |\mu| \delta r \hat{N}_+} = e^{-i |\mu| \delta r } \hat{c}_+ \,.
\end{equation}
Consequently, the squeezing parameter evolves as
\begin{equation}
    \xi \longrightarrow \Tilde{\xi}(\delta r) = r_s e^{i(\vartheta - 2|\mu| \delta r)} \,,
\end{equation}
and the evolved state, keeping only the relevant generator $\hat{H}_I$, can be written as 
\begin{equation}
    \ket{\psi_{\mathrm{ss}}(\delta r)} = \hat{S}_+ (\Tilde{\xi}) \ket{0} \,.
\end{equation}
Therefore, the parameter is encoded entirely in the orientation of the squeezing ellipse, $\vartheta - 2|\mu| \delta r$. 

\begin{figure}[t!]
\centering
\vspace{5mm}
\includegraphics[width=0.99\linewidth]{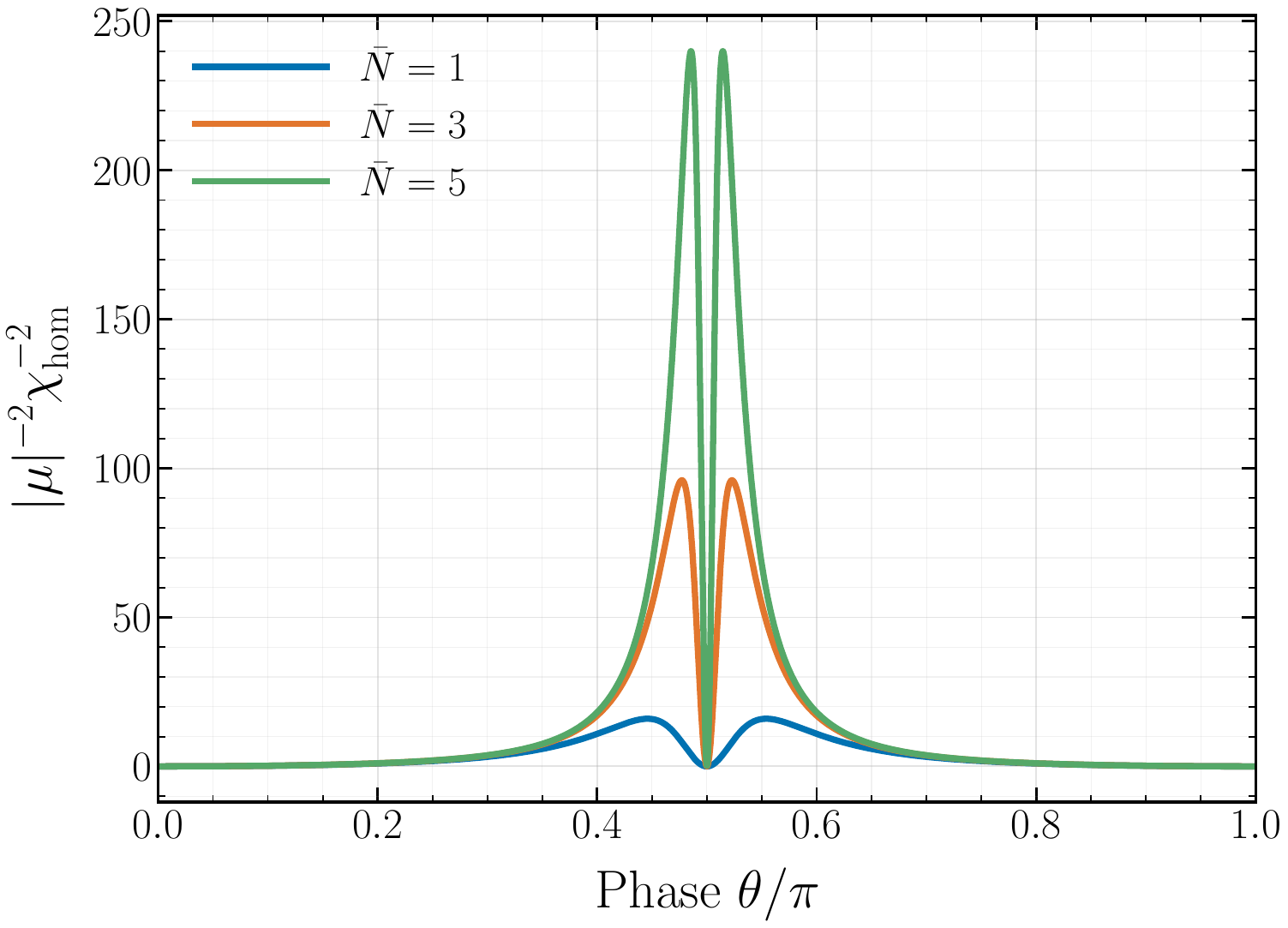}
\caption{The moment-based sensitivity for the homodyne detection against the local oscillator phase $\theta$, for different average photon numbers and squeezing phase $\vartheta=\pi$.}
\label{Fig: Homodyne detection}
\end{figure}

The expectation value and its derivative can be straightforwardly computed
\begin{align}
    &\langle \hat{X}_\theta^2\rangle = \frac12 \bigg[ \cosh(2r_s) - \sinh(2r_s) \cos(2\theta - \vartheta + 2|\mu|\delta r) \bigg] \,, \\
    &\frac{\partial}{\partial\delta r}\langle \hat{X}_\theta^2\rangle \Bigg|_{\delta r =0} = |\mu| \sinh(2r_s) \sin(2\theta - \vartheta) \,.
\end{align}
Note that the sensitivity therefore vanishes when $2\theta = \vartheta$, that is when the local oscillator is aligned exactly with the squeezed principal axis. We can see how the moment-based sensitivities vary with changes in phase difference in Fig.~\ref{Fig: Homodyne detection}. The two maxima correspond to the optimal phase $\theta_{\mathrm{opt}}$, while the two minima, for which the sensitivity vanishes, correspond to $2\theta - \vartheta = 0$ or $\pi$. The two maxima are extremely close to one of the minima, and the performance becomes highly sensitive to shifts in $\theta_{\mathrm{opt}}$ in an asymmetric way for high average photon numbers. 

To find the variance, we first note that for our centred Gaussian \cite{Weedbrook:12}
\begin{equation}
    \langle \hat{X}_\theta^4\rangle = 3 \langle \hat{X}_\theta^2\rangle^2 \,.
\end{equation}
Thus,
\begin{equation}
    \Delta^2 \hat{X}_\theta^2 = 2 \langle \hat{X}_\theta^2\rangle^2 \,,
\end{equation}
and the sensitivity $\chi^{-2}_{\mathrm{hom}} := \chi^{-2} [\hat{\rho}_{\mathrm{ss}}, \hat{X}_\theta^2]$ at $\delta r =0$ is given by
\begin{equation}
    \chi^{-2}_{\mathrm{hom}} (\theta) = \frac{2 |\mu|^2 \sinh^2(2r_s) \sin^2(2\theta-\vartheta)}{\big( \cosh(2r_s) - \sinh(2r_s) \cos(2\theta - \vartheta)\big)^2} \,.
\end{equation}
A straightforward optimisation of the above expression reveals that the optimal homodyne angle is
\begin{equation}
    \cos(2\theta_{\mathrm{opt}} - \vartheta) = \tanh(2r_s) \,.
\end{equation}
For this optimal angle, we obtain the sensitivity 
\begin{equation}
    \chi^{-2}_{\mathrm{hom}} (\theta_{\mathrm{opt}}) = 2|\mu|^2 \sinh^2(2r_s) = 8|\mu|^2 \Bar{N}(\Bar{N}+1) \,.
\end{equation}
This saturates the leading-order term of the corresponding QFI, $F_{Q}^{\mathrm{HG}} [\hat{\rho}_{\mathrm{sq}} ]$ in Eq.~\eqref{eq: Fq max HG sq}, while missing the contributions from the orthogonal vacuum modes into which displacement causes leakage. This is because the coupling to these inaccessible modes modifies the accessible quadrature statistics only at second order in $\delta r$. If we assume that the experiment only has access to the originally populated modes, we would necessarily need a measurement observable that is singular at the working point, i.e. $\Delta^2 \hat{M}\big|_{\delta r=0} = 0$, for the given initial state. This was possible for the parity measurements considered in the previous subsections. For the squeezed-vacuum state, however, we could not construct a parity-based observable whose moment-based sensitivity saturates the full QFI. This is not to say that no theoretical observable could achieve this. For example, a projection onto the original squeezed vacuum state, $\hat{M} = \hat{\rho}_{\mathrm{ss}}$, would give the locally optimal sensitivity, but this is not operationally trivial in quantum optics \cite{Frascella:19}. 

Compared with the twin-Fock state sensitivity (for which $\Bar{N}=N$) with parity measurement, we see that the squeezed state with homodyne measurement performs a factor of 4 better in the quadratic term in the best-case scenario. The experimental challenges for the two measurement schemes also differ. Creating balanced twin-Fock states for large photon numbers is not trivial. Nonetheless, average photon numbers comparable to the squeezed states may already be reached \cite{Sturges2021}. On the other hand, squeezed vacuum states can be routinely prepared in the lab, but maintaining the phase $\theta_{\mathrm{opt}}$ is challenging. However, it is not impossible and active phase-locking techniques have demonstrated sophisticated control of the local oscillator phase \cite{Wang2019}.

\section{Conclusions}

We investigate the estimation of the magnitude of transverse beam displacement for fixed photon number states and fixed average photon number pure Gaussian states, with Hermite-Gauss and Laguerre-Gauss modes, using a novel theory for mode parameter estimation. Restricting ourselves to just a two-mode setup in the limit of small displacement and no photon losses, we find the optimal initial state and the corresponding quantum Fisher information for scenarios in which the total number of photons is fixed. This gives a theoretical upper bound on the sensitivity for estimating the displacement for all lossless, small-displacement, two-mode, fixed-$N$ setups. It scales quadratically in the photon number and thus demonstrates the possibility for the sensitivity to reach Heisenberg scaling. The initial state asymptotically approaching the optimal state for large photon numbers is found to be a rotated NOON state. 

Conversely, we also look at two-mode centred pure Gaussian states, and compute the quantum Fisher information for the state that maximises the variance of the effective subspace Hamiltonian. This state is a single-mode squeezed state in a rotated basis and is asymptotically optimal for all two-mode Gaussians. We show that this Gaussian provides an improvement in sensitivity over the theoretical limit for the fixed-$N$ states by a factor of around 2. This is due to the photon number fluctuations of the Gaussian states without the finite photon budget constraint. 

Finally, we propose concrete measurement schemes using the method of moments and calculate their moment-based sensitivities. We propose approaches reaching quantum-enhanced scaling for both fixed-$N$ and squeezed vacuum states. 

Our work shows that reaching the Heisenberg scaling in the sensitivity for estimating the magnitude of transverse beam displacement is possible for certain setups with bounded photon numbers, and it establishes ultimate limits and their corresponding optimal initial states. While we focused on Hermite-Gauss and Laguerre-Gauss modes due to their commonality in the literature, other mode bases could also be considered. The methods can be readily extended to three or more modes, potentially improving the sensitivities. Furthermore, a similar analysis could be carried out on the estimation of the beam displacement direction. Considering both the displacement magnitude and direction would lead to a multi-parameter estimation problem, which requires further study. An additional enhancement could also be found using exotic quantum states with constraints different from those considered in this paper. Future work could explore robustness to loss and noise, including comparisons with squeezed-state strategies. Optimisation under these conditions may also reveal new optimal states at finite photon number.

\textit{{Acknowledgements---}} 
This work was funded by the project PID2023-152724NA-I00, with funding from MCIU/AEI/10.13039/501100011033 and FSE+, by the project CNS2024-154818 with funding by MICIU/AEI /10.13039/501100011033, by the project CIPROM/2022/66 with funding by the Generalitat Valenciana, and by the CSIC Interdisciplinary Thematic Platform (PTI+) on Quantum Technologies (PTI-QTEP+). This work is supported through the project CEX2023-001292-S funded by MCIU/AEI. 

M. Gessner and P. Colomer gratefully thank the support provided by the ``Institut de Ciències Fotòniques'' (ICFO) in the early stages of this project.

P. Colomer acknowledges the financial support by the German Federal Ministry of Research, Technology, and Space (BMFTR) within the program “Souver\"an. Digital. Vernetzt.” – Joint project 6G-life 16KIS2414 for TUM; project QUARKS 16KIS1998K; project Q-TREX 16KISR027; and project QUIET 16KISQ092.

\textit{{AI usage statement---}} 
We acknowledge the usage of AI-based tools to assist with language editing, clarity enhancement and verify intermediate derivations during manuscript preparation. All scientific analyses, research findings, and conclusions were developed and validated by the authors, who assume full responsibility for the content of this work.

\bibliography{apssamp_formatted}

\end{document}